# Study of Silicon Pixel Sensors for Synchrotron Radiation Detection

LI Zhen-Jie(李贞杰)[1,2] JIA Yun-Cong(贾云丛)[2] HU Ling-Fei(胡凌飞)[1] LIU Peng(刘鹏)[1]
YIN Hua-Xiang(殷华湘)*[2]

[1] Institute of High Energy Physics, Chinese Academy of Sciences, Beijing 100049, China

[2] Institute of Microelectronics of Chinese Academy of Sciences, Beijing 100029, China

*yinhuaxiang@ime.ac.cn

**Abstract:** Hybrid pixel single-photon-counting detectors have been successfully employed and widely used in Synchrotron radiation X-ray detection. In this paper, the silicon pixel sensors for single X-ray photon detection, which operate in full-depletion mode have been studied. The pixel sensors were fabricated on 4-inch, N type, 320μm thick, high-resistivity silicon wafers. The pixel sensors has a p+-in-n structure with varies of pixel size and gap size including guard-ring structures. Later, the pixel sensor was wire bonded to the ASIC circuits and tested for the performance of X-ray response in the synchrotron beam line (BSRF, 1W2B). From the S-curve scan, we could get the energy resolution and the linear properties between input energy and the equivalent generator amplitude. The pixel sensors we fabricated have a good energy linear and high count rate depending on the ASIC readout circuit. We get the 20% energy resolution above 10 keV photon energy via wire bonding. The energy resolution would get better if we bond the sensor via indium ball for the smaller stray capacitance.

**Key words:** synchrotron X-ray, silicon pixel sensor, dark current, energy resolution, count rate

**PACS:** 29.40.Wk, 29.20.dk

## 1. Introduction

For the low dark current noise, short readout time in millisecond range, high dynamic ranges, hybrid pixel single-photon-counting detectors have been successfully employed in Synchrotron radiation X-ray detection [1-2]. The pixel sensor is the main element of the pixel detector as well as the readout electronics which bump-bonding to the pixel sensors. To design and fabricate the sensor, the semiconductor planar process is used, including oxidation, etching, implanting and sputtering [3]. For the considerations of low dark current, high breakdown voltage and suitable total depletion voltage, optimization simulation should be taken for some special structures such as guard rings, inner pixels gaps [4-5]. Of course, the technical process should be optimized by TCAD to reduce the oxide trapped charge between field oxide and silicon, decrease the junction dead depth and optimize the electric field.

The purpose of this paper is to study the pixel sensors for synchrotron radiation detection. The silicon pixel sensor will be applied in the pixel array detector for the BAPS (Beijing Advanced Photon Source).The pixel dark current, reverse junction capacitance have been tested using probe station. Later, the sensor's response to X-ray radiation has been tested via bonding the sensor to the ASIC circuit which was specially designed for the silicon pixel detector. From the ASIC S-curve, we calculated the sensor's energy resolution and the linear response between input X-ray energy and the equivalent generator amplitude. In this paper, the pixel sensor design and process technology were firstly introduced. Then we simulated the important process and the sensor device electrical properties. Finally, we bonded the sensor to the ASIC and test the response to X-ray radiation.

## 2. Overview of the Pixel Sensors Flow

The pixel sensor should be designed carefully to reduce the dark current and to avoid the breakdown [6]. The technology process should also be optimized to decide the process parameters for suitable oxide thickness, implant dose and energy, junction thickness [3]. We have design varies of pixels to simulate and test the characteristics. Pixels with different size, gap and different multi-guard-rings have been simulated for their electrical properties. The complete assembly consists of several arrays of test structures with 8x8 pixels and some single diodes on the wafer. Pixel arrays were made on a 4-inch-diameter *n*-type <111> wafer with a thickness of 300μm and resistivity of 6-8 $k\Omega\cdot cm$. The pixel size ranges from 180μm to 150μm with different considerations. The typical parameters of the p+-in-n pixel sensors are summarized in Table 1.

The pixel sensor was designed to square shape with five guard rings. Figure 1 shows a two dimensional basic drawing of the device. The 8x8 pixels were used to test the dark current and junction capacitance. In the semiconductor industry, technology CAD tool has been used to optimize the sensor structure and verify the fabrication process. Though the TCAD is capable of 3D simulations, we use the TCAD to simulate the sensor in two dimensions, which resulted in simplicity that was sufficient to elucidate the sensor properties we explored. The processing step items of the TCAD are as follows: oxidation, deposition, etching, ion implantation, and annealing.

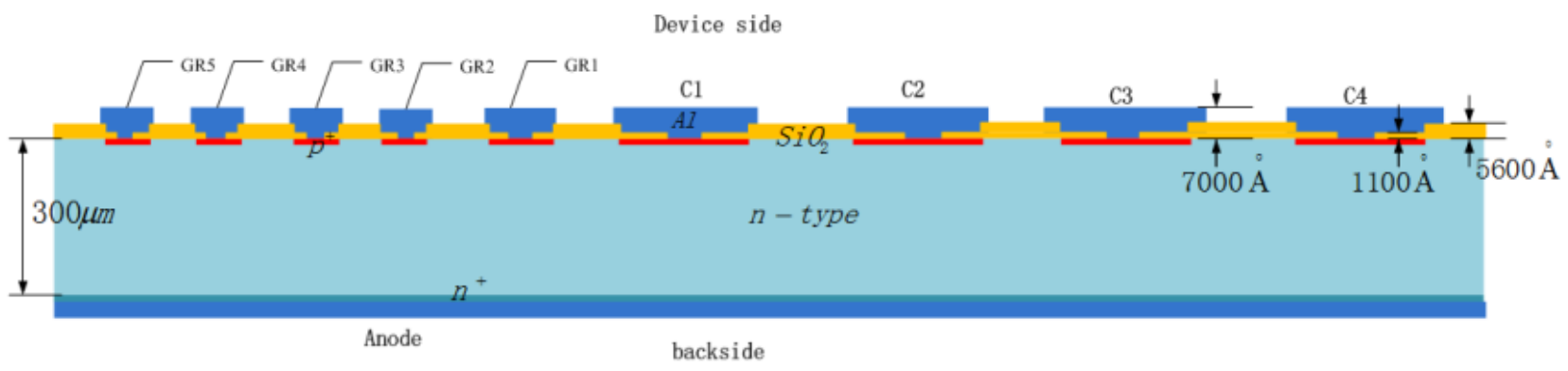

Fig.1.The sketch drawing of pixel sensor with five guard rings on the wafer of 300um thickness

The processing on the device side is as following: the original 5600 Å of silicon dioxide grew through wet oxide. This step should be controlled carefully to get a good quality oxide at the Si-$SiO_2$ interface whereas the bulk oxide was grown through wet process to reduce time for growth of a 0.56μm thick oxide. First the lithography was used to generate the $p^+$ area by removing the 5600 Å oxide. Then the gate-oxide was generated on the $p^+$ area. A 35 keV boron implantation through silicon dioxide was used to generate the $p^+$ pixels. Wet etching was then performed to remove the 1060 Å silicon dioxide layer which covered the metal contact locations. On the entrance window side, the 110 keV phosphorus implantation was used to generate the ohmic contact without gate-oxide. The implantation was followed by an annealing step for about 30 minutes. In order to make contact with metal, the contact windows are performed in the next step. Both the device side and entrance side, a 7000 Å thick aluminum layer was deposited, the aluminum film was then patterned by a lithography step. The entrance window side is radiation hard due to the lack of silicon dioxide between the heavy doping and the aluminum layer. These processes require only three masks and no photolithographic steps are applied on the backside. Figure 3 shows the finished wafer without passivation layer.

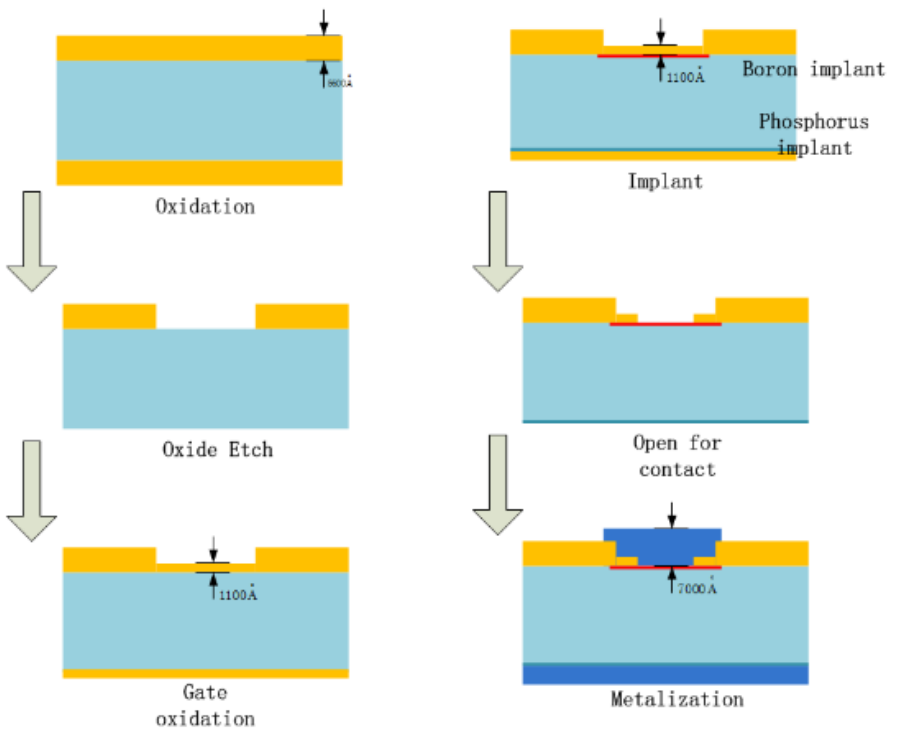

Fig.2. Flow process chart

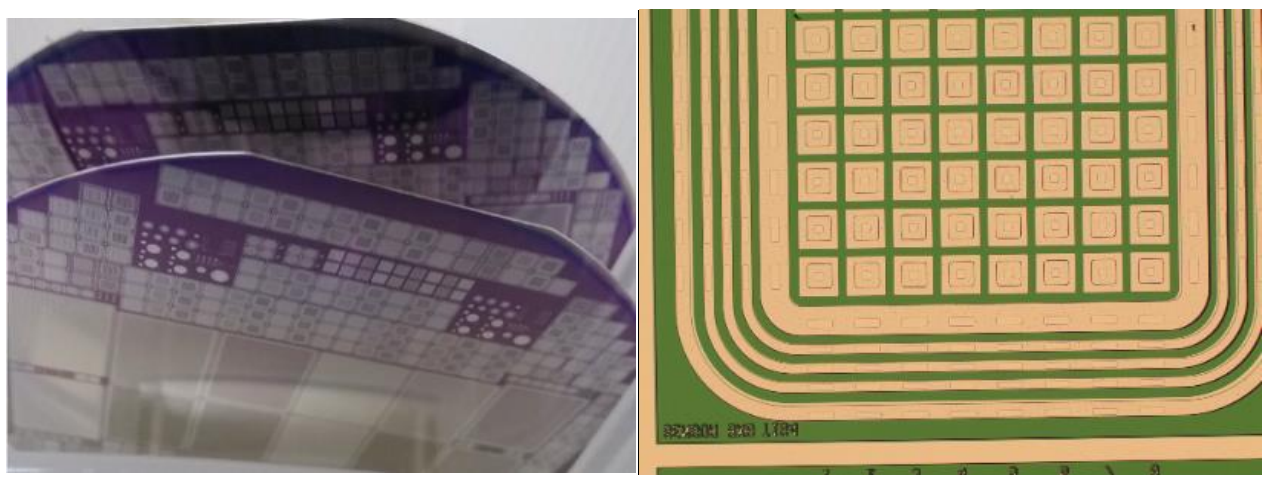

Fig.3. Image of finished wafers

**Table 1**

Typical parameters of $p^+$-in-n pixel sensors.

| | | |
|---|---|---|
| Silicon wafer | 4 in. n-type | |
| Crystal | <111> | |
| Resistivity | 7 kΩcm | |
| Thickness | 300-320μm | |
| Pixel size | 180×180μm$^2$ | 150×150μm$^2$ |
| Number of Pixels | 8×8=64 for test | |

## 3. Optimization of Pixel Sensors

The pixel sensor is designed as a single side planar structure with PIN diodes having p+-implantations in an n-type substrate. The pixels are the signal collecting electrode (cathode) applied with ground potential, and the backside (anode) is applied with positive full depletion voltage as sketched in figure 1. That leads to the electrons drift to the anodes and the holes drift to the cathodes. We simulated the depth of implant junction to determine the dead layer. Because the junction area could not act as the absorb region but as the collected charge region. The guard rings area was also simulated to understand the potential distribution and to avoid breakdown for the case of high voltage applied on the sensor. The full depletion voltage was one of the important parameters. For the pixel sensor should work above the full depletion voltage for the maximum quantum efficiency.

### 3.1 The Junction Depth with Implant Energy

The junction region forms the dead region for X-ray detection. In order to minimize the dead region, shallow implant and junction should be used in pixel sensors. To determine the junction depth, we simulate the implantation process using the TCAD. Fig 4 shows the junction depth with different implantation energy. The Boron implant on the device side forms the $p^+$-$n$ junction with the $n$-type substrate. In consideration of the junction depth and technology limit, we choose the implant energy of 35 keV, which could make the junction depth about 1μm. The backside implant of phosphorus was to form the ohmic contact with the aluminum electrode. In the simulation process, the initial 1060 Å oxide was also used to serve as the implant stop layer to decrease the substrate's damage.

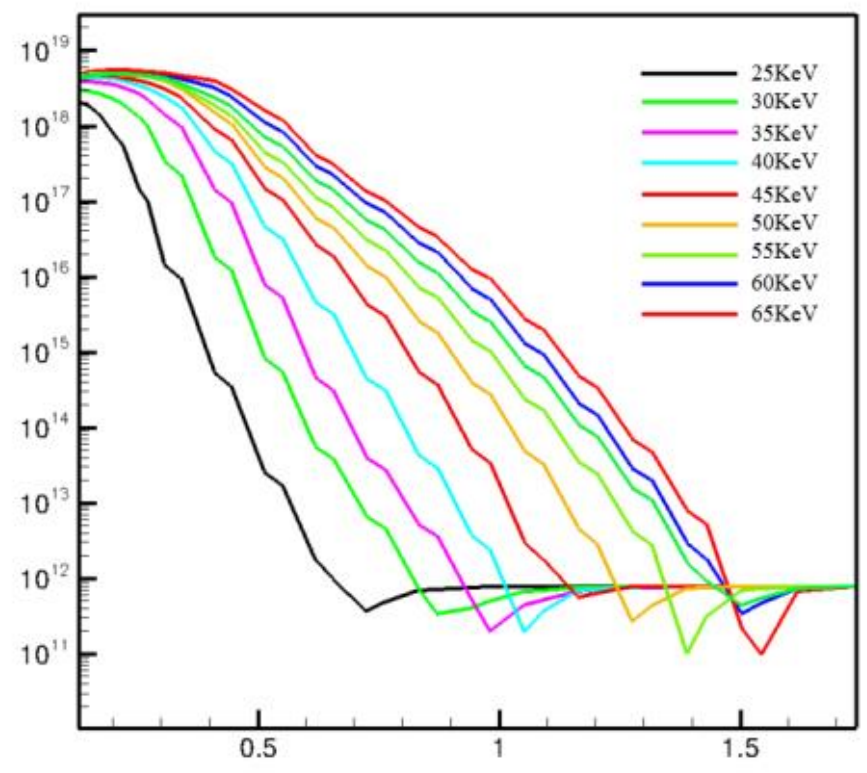

Fig.4.The Boron junction depth at different implant energies with the same dose (dose=$6e^{14}/cm^3$)

and 730 °C , 30 minutes annealing

### 3.2 Design of the Guard Ring Area

The multi-guard ring structure is to form a gradual voltage drop between the sensitive region on ground potential and the cutting edge on backside potential [7-8]. This kind of structure is to reduce the dark current in the active region by decoupling the current generated outside the active region and prevents the space charge region from reaching the cutting damage area. It can also eliminate high field region that could cause the avalanche breakdown [7-10]. The guard rings could be applied with ground potential or serve as floating rings and they bias themselves via the punch- through-mechanism. The number of the guard rings and the width of the gaps are determined by the maximal applied voltage, and the thickness of the sensor. With optimal guard ring number and gap width, could the dead region be minished. Figure 5 shows the potential of the pixel sensor with five guard rings floating and the anode being biased at 100*V*. The potential drops slowly from the left guard ring to the pixel area.

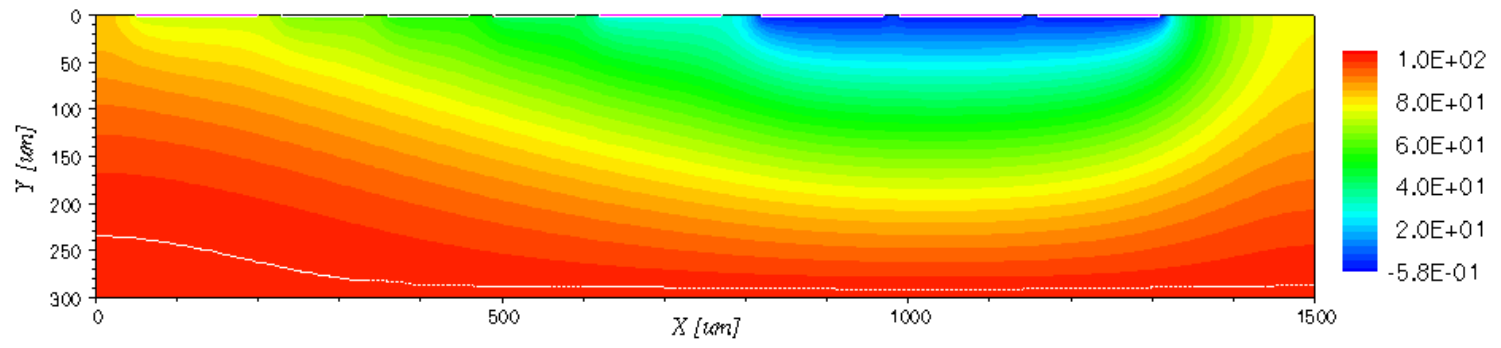

(a)

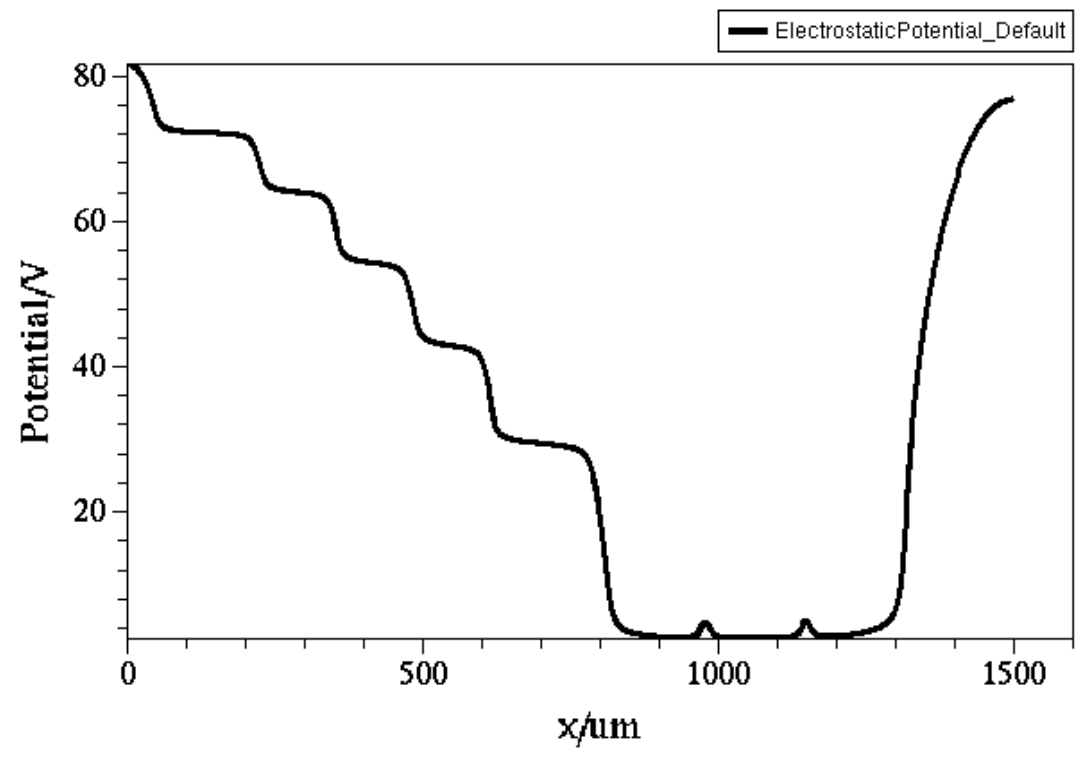

(b)

Fig.5. (a) The potential distribution of pixels sensor with guard rings floating and the backside applied with 100*V*, device side with 0*V*, (b) the x direction potential distribution at y=20μm

### 3.3 Heavy Ion Pulse Response with Different Bias Voltage

The pixel detector was used to detect single synchrotron X-ray photons, and X-ray photons will produce electron-hole pairs once the photons were absorbed in the silicon body. For a short wave length of X-ray, the simulations of X-ray response become difficult for the TCAD. Usually the TCAD simulates long wave length photons such as visible light and infrared light. But, the TCAD provides radiation models for the possible simulations of injected electron-hole pairs. When the high-energy particles penetrate a semiconductor device, they deposit their energy by the generation of electron-hole pairs just like the X-ray photons. These charges can move along the electric field lines and induce current, and then will be collected by the electrode at the end. We simulate the Heavy Ion response of silicon device, and the induced current shows the process of charge collection. These simulations will help the designs of silicon device and the ASIC readout electronics. Figure 6 (a) shows the current of holes and the 2D current density map at different time points, they show the electron-hole pairs collection process. The charge cloud spreads when it moves along the electrical field and this diffusion will cause the effect of charge sharing especially in small pixels. The total induced current and charge vary with the time, shown in Figure 6 (b). We can see that the total current decreases to zero at about 12ns with voltage equal to 100V, so the signal will be totally collected within 12ns. The bigger bias voltage, the faster collected the charge. But the bias voltage should not exceed the breakdown voltage.

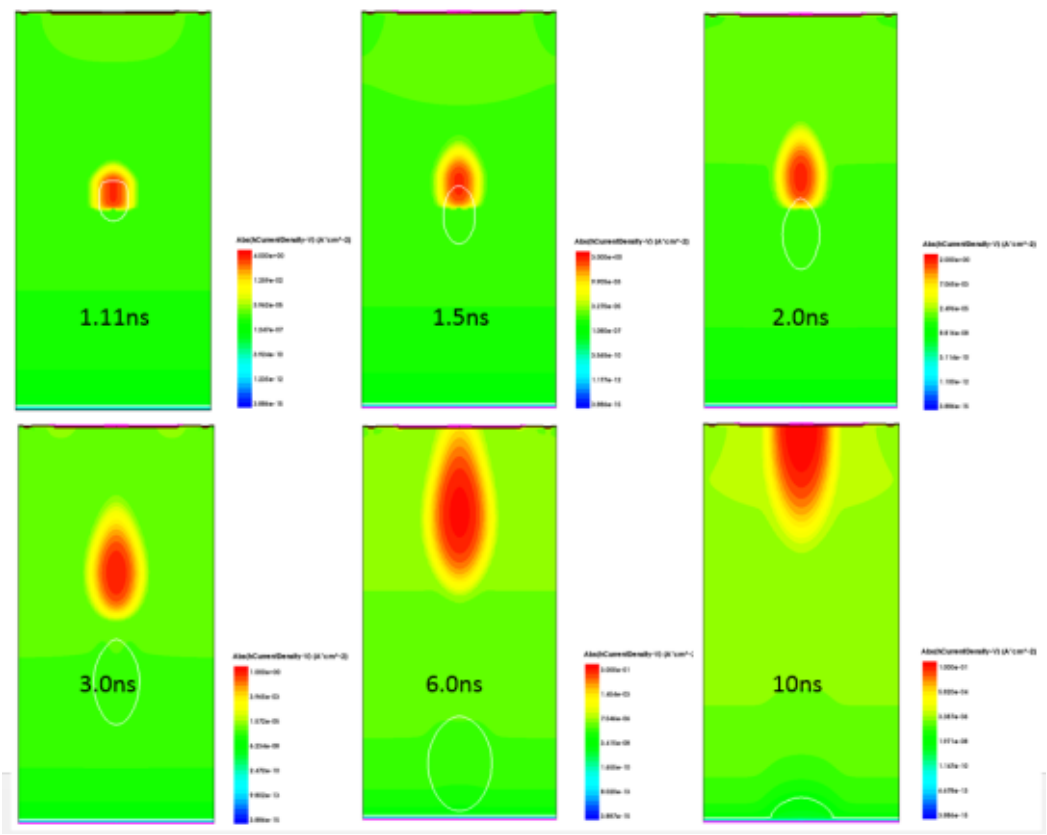

(a)

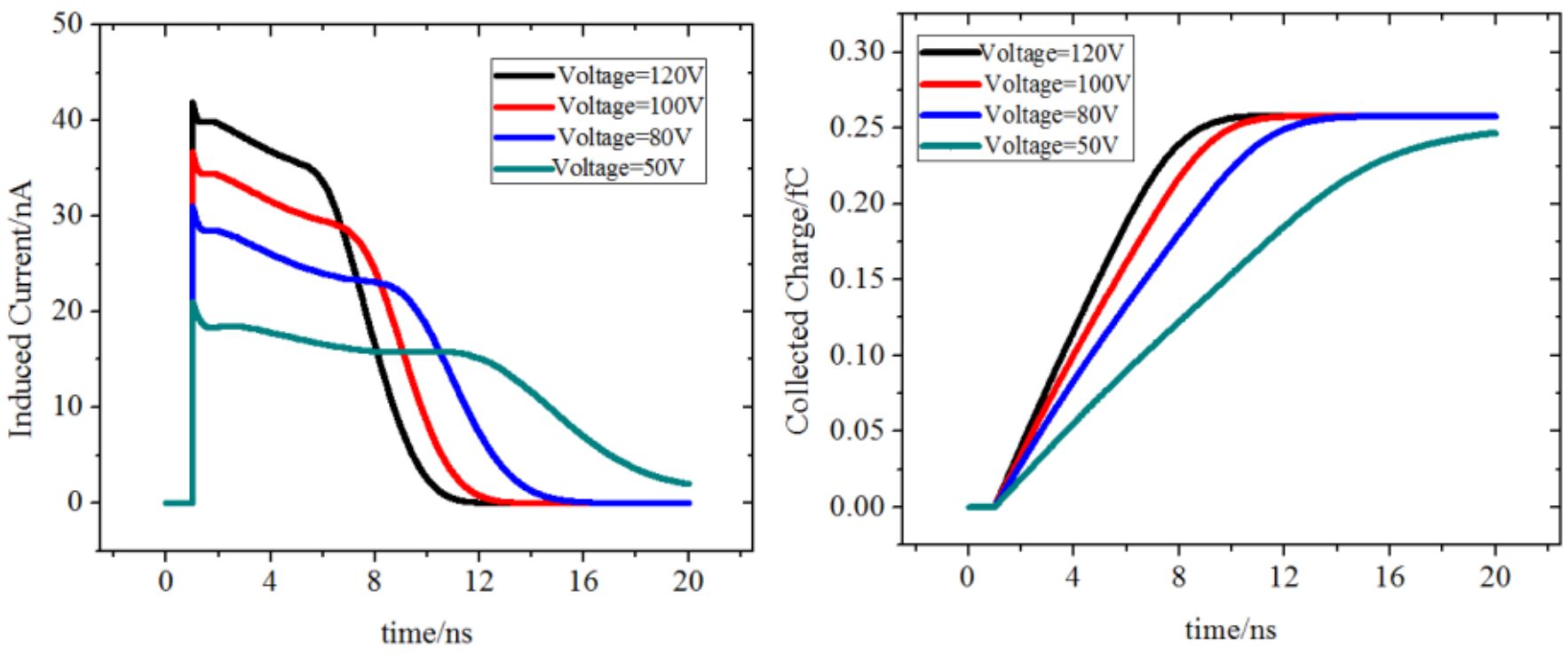

(b)

Fig.6. (a) The holes drift from the initial position to the electrode at different time points, (b) The induced current in electrode and charge collection at different bias voltages.

## 4. Characterization of the Electrical Properties and Synchrotron Radiation Response

After the device was finished, we tested the electrical properties including full depletion capacitance of pixels and the total leakage current of different structure pixels. Then, we bonded the sensor to ASIC circuit to test the response to X-ray radiation on synchrotron radiation facility.

### 4.1 C-V Measurements and Fully Depletion Voltage

In order to collect full charge in the detector, the silicon sensor has to be depleted of free and mobile carries by applying $V_{bias}>V_{FD}$ in reverse direction. The bulk capacitance per unit area is determined by the depth of depletion layer *w*,

$$C_{bulk} = \frac{\varepsilon_{Si}}{w} = \sqrt{\frac{\varepsilon}{2\mu\rho V_{bias}}} \qquad (4)$$

Where $\mu$ is the mobility and ρ is the resistivity. When the bulk is fully depleted $w=D$, where $D$ is the thickness of the substrate. Formula (4) resolves to

$$V_{FD} = \frac{D^2}{2\varepsilon_{Si}\rho\mu} \qquad (5)$$

After depleting the whole bulk, the capacitance $C_{bulk}$ will remain constant. Fig 8 shows the capacitances as a function of the reverse bias voltages, the measurement shows a good agreement in the fully depletion voltage below 50V between the fabricated sensors and simulation result.

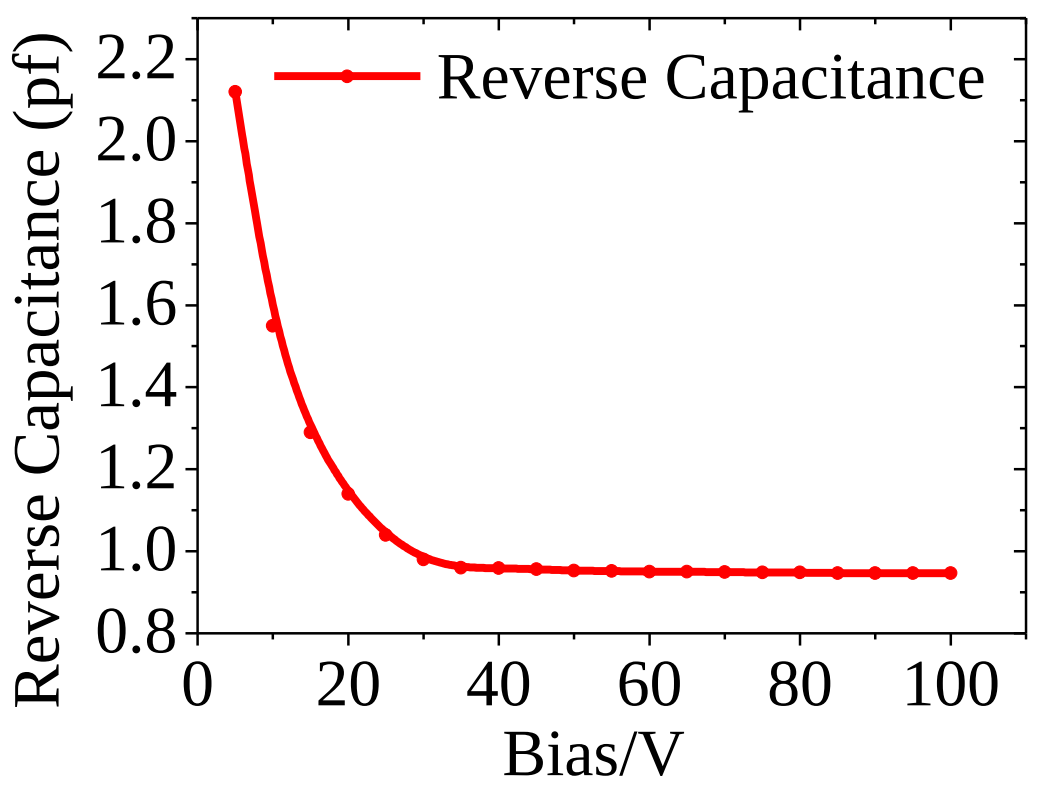

Fig.8. The reverse capacitance of p-n junction----the reverse capacitance gets minimum when the bias voltage reached to the value of fully depletion voltage

### 4.2 I-V Measurements and Dark Current

The dark current is the major criteria for the quality of the silicon sensor. The leakage current and total capacitance counting in the contact capacitance and bulk capacitance are the major noise source of readout system, which should be kept as low as possible. Of course, the bulk capacitance is constant when $V_{bias} \geq V_{FD}$, which would not be decreased anymore and the contact capacitance could be decreased by indium bump bonding. The leakage current is proportional to the depletion layer thickness *d* of the sensor, which is proportional to $\sqrt{V_{bias}}$. So it should be constant after $V_{bias}$ reaching $V_{FD}$ before the sensor approaching breakdown. We test single pixel leakage current of T1S1~T1S4, T2S2~T2S4, T3S3, T4S1 structures and the 8x8 pixels leakage current of T1S1. The test results (figure 8) show the sensor's performance and a typical leakage current level of 0.1nA/pixel, 0.5nA/64pixels without guard rings grounded, 0.05nA/64pixels with guard rings grounded. From the test result above, we can conclude that the developed silicon pixel sensor's electrical characteristics satisfy the goal of 1nA/cm$^2$ and below 50$V$ full depletion voltage and above 150$V$ breakdown voltage.

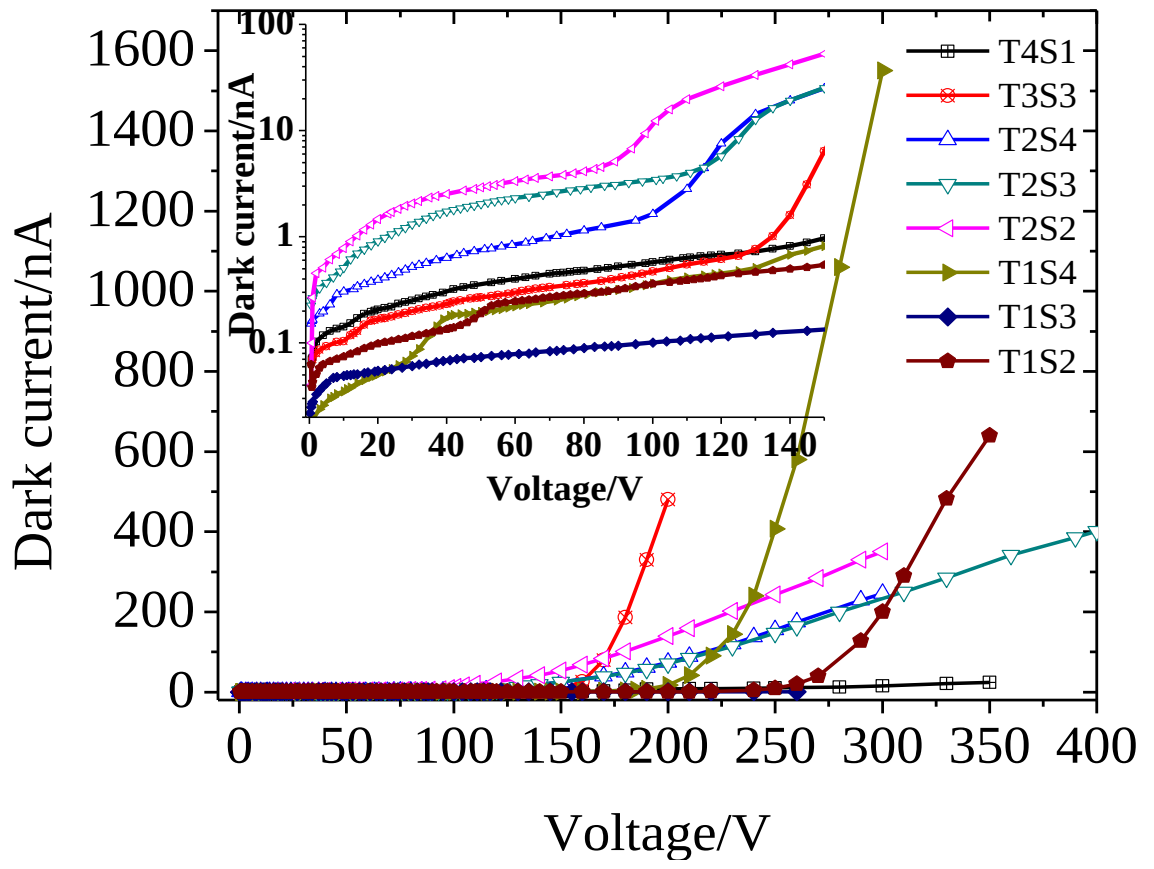

(a)

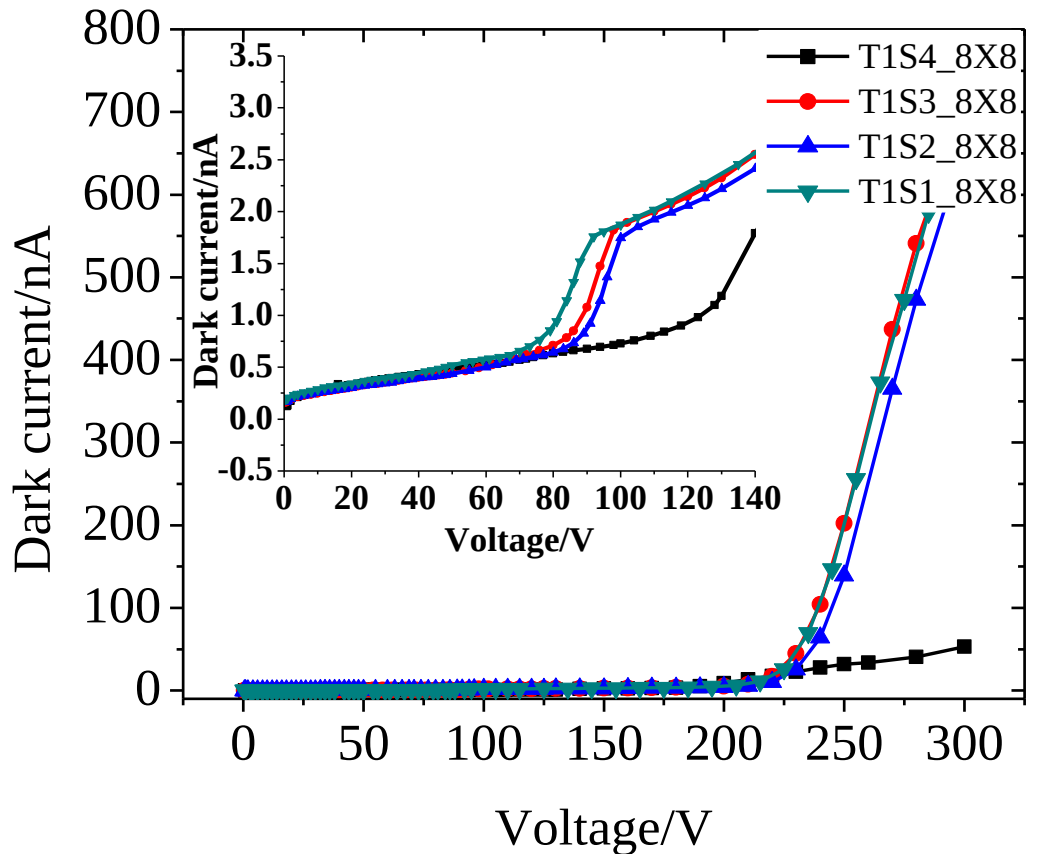

(b)

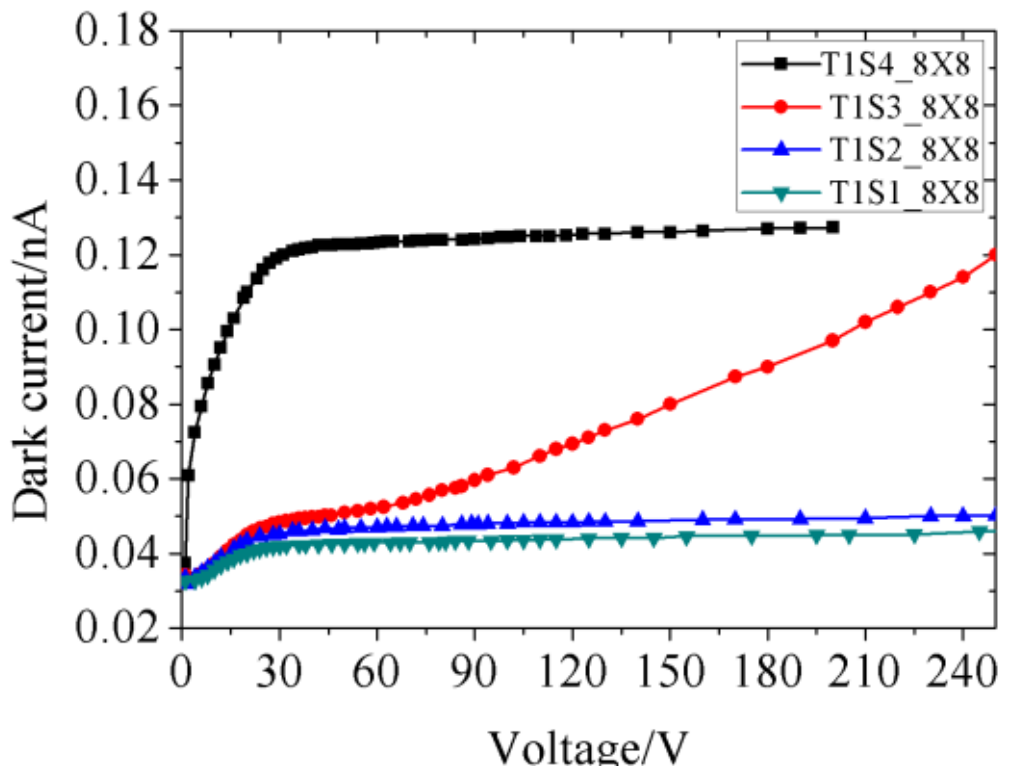

(c)

Fig.9. The dark currents are plotted as a function of bias voltage (a) Single pixel leakage current of *T1 T2 T3 T4* series without guard rings grounded, (b) 8x8 pixels leakage current of *T1* without guard rings grounded and (c) with guard rings grounded

The leakage current was different from the outer pixels to inner pixels because of the cut edge of sensor. The outer pixels leakage current would be larger than the inner's generally. Figure 10 shows the distribution of 8×8 pixels leakage current, it's clear the leakage current of inner pixels is smaller than that of the outer.

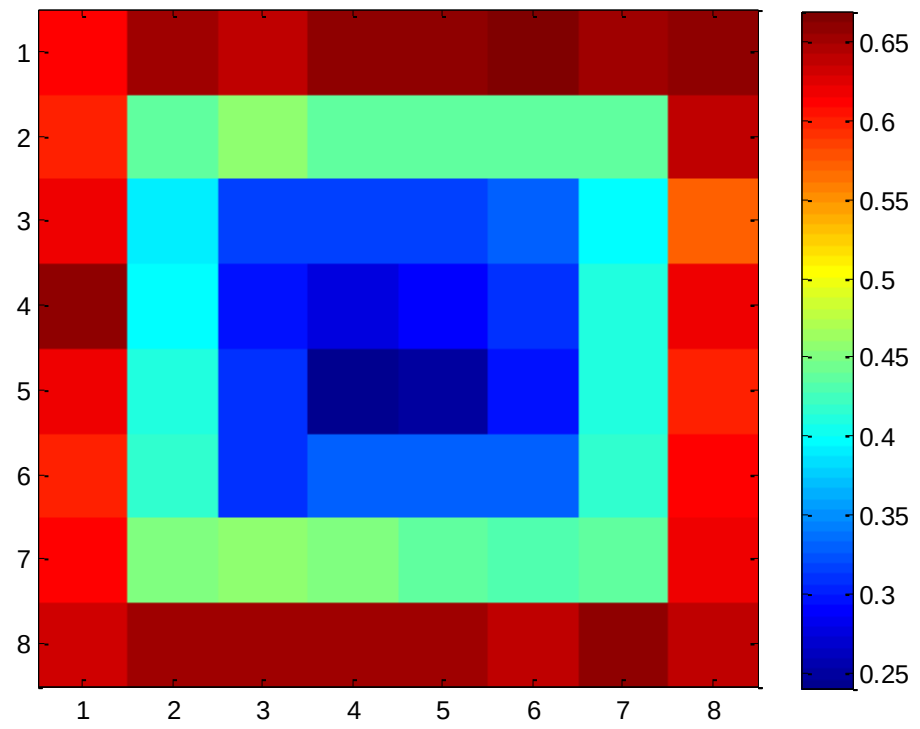

Fig.10. 8x8 pixels leakage current distribution (unit/nA), the outer pixels' dark currents are larger than that of the inner's.

## 4.3 Synchrotron Radiation Test

The sensor was bonded to the ASIC circuit by wires, figure 11(a).For simplification, we connected four pixels to the ASIC readout circuit and tested the X-ray response. Figure 11(a) shows the details of the pixels, the inner guard ring was also bonded to the circuit. In the synchrotron beam line (BSRF, 1W2B), the X-ray energy could varies from 8keV to 18keV that was the energy we used. Figure 11(b) shows the experimental facility for our test, the collimation hole is used for the X-ray collimation and we can change the X-ray intensity by rotate the dial.

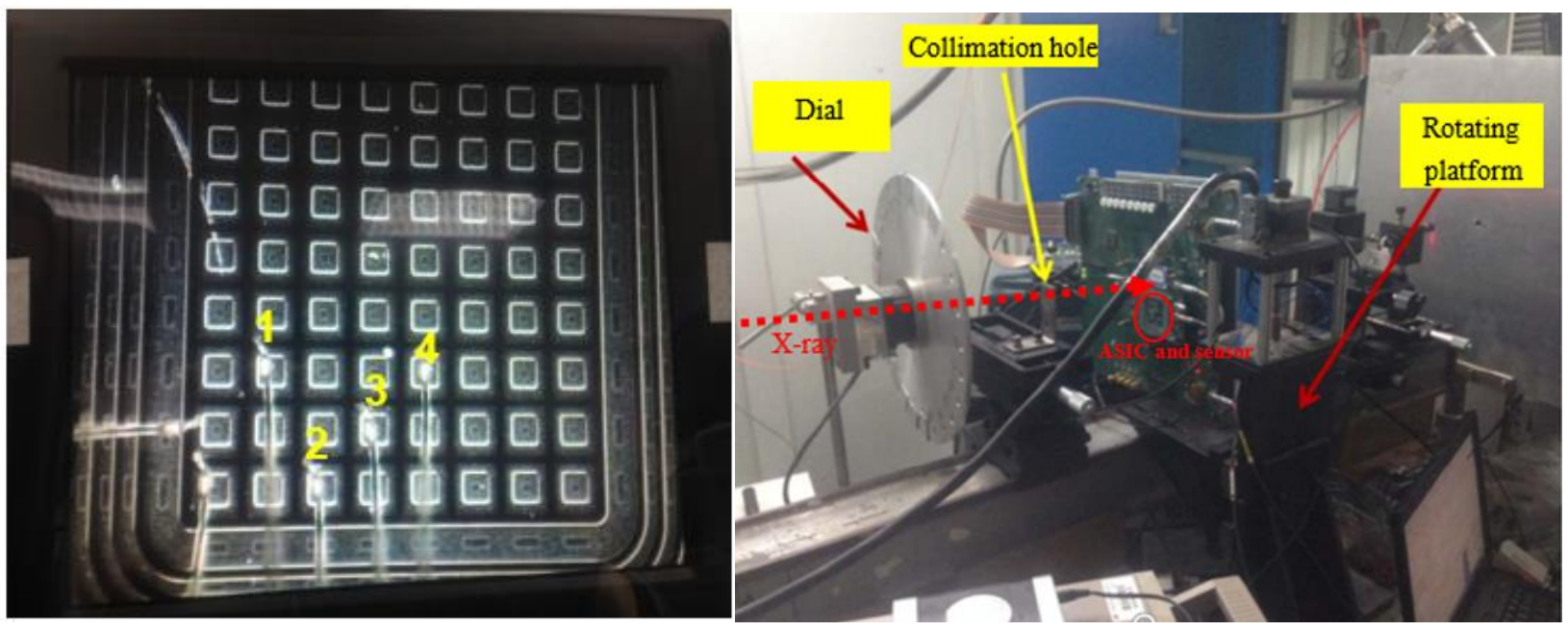

(a) (b)

Fig.11. (a) The picture of 8x8 pixels sensor with four pixels wire bonding to ASIC, (b) the facility of synchrotron X-ray test, the X-ray beam irradiate on the junction side of the sensor

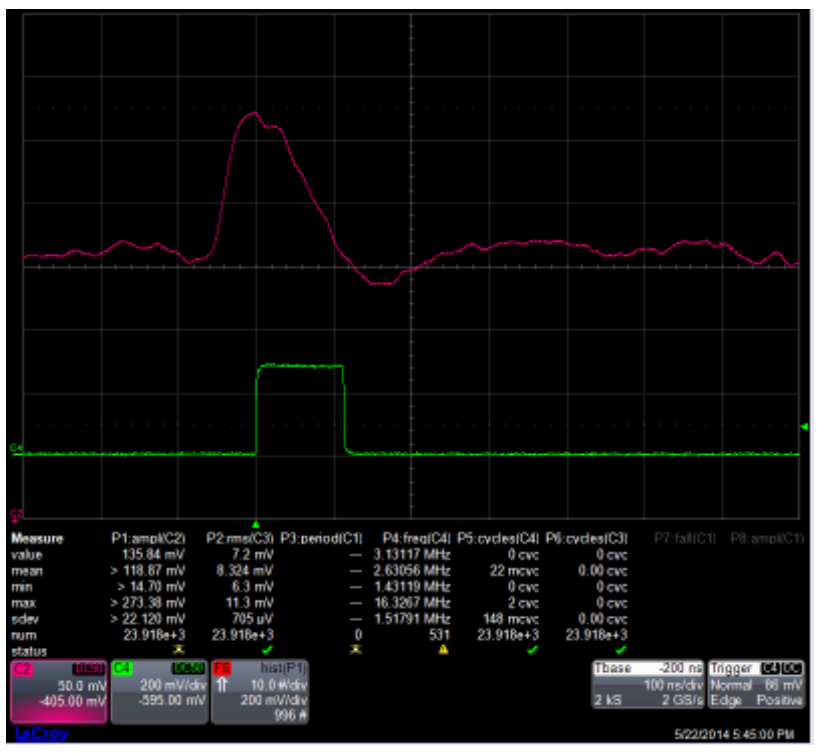

Fig.12. the analog output signal of ASIC (red line) and the discriminator output (green line)

The ASIC includes charge sensitive preamplifier, shaping circuit, threshold discriminate circuit. Figure 12 shows the signal of 8KeV photon, the upper signal (red line) is the output of charge sensitive preamplifier. The ASIC could count when the signal’s amplitude is larger than the threshold discriminate amplitude whose value is set via Equivalent Generator, such as the nether signal (green line) of figure 12. The signal amplitude is different for different X-ray photon energies. For a fixed X-ray energy, the signal from the shaping circuit will be constant. So when the threshold amplitude is larger than the signal, the counting is zero, otherwise the counting is the number of X-ray photons. Fig 13 shows the count performance of pixel sensor at different X-ray energy. The four pixels show the same trend for the photons counting. the photon number of the pixel 1 count maximum and other three pixels’ are less because of the distribution of the light spot. Figure 14(a) shows the performance of pixel 1 at different energies, and we can see that the decrease points move as the energies change. The equivalent generator amplitude at decreased points should be proportional to X-ray energy, just as figure 14(b), which shows a good linear between the equivalent generator amplitude and the X-ray energy.

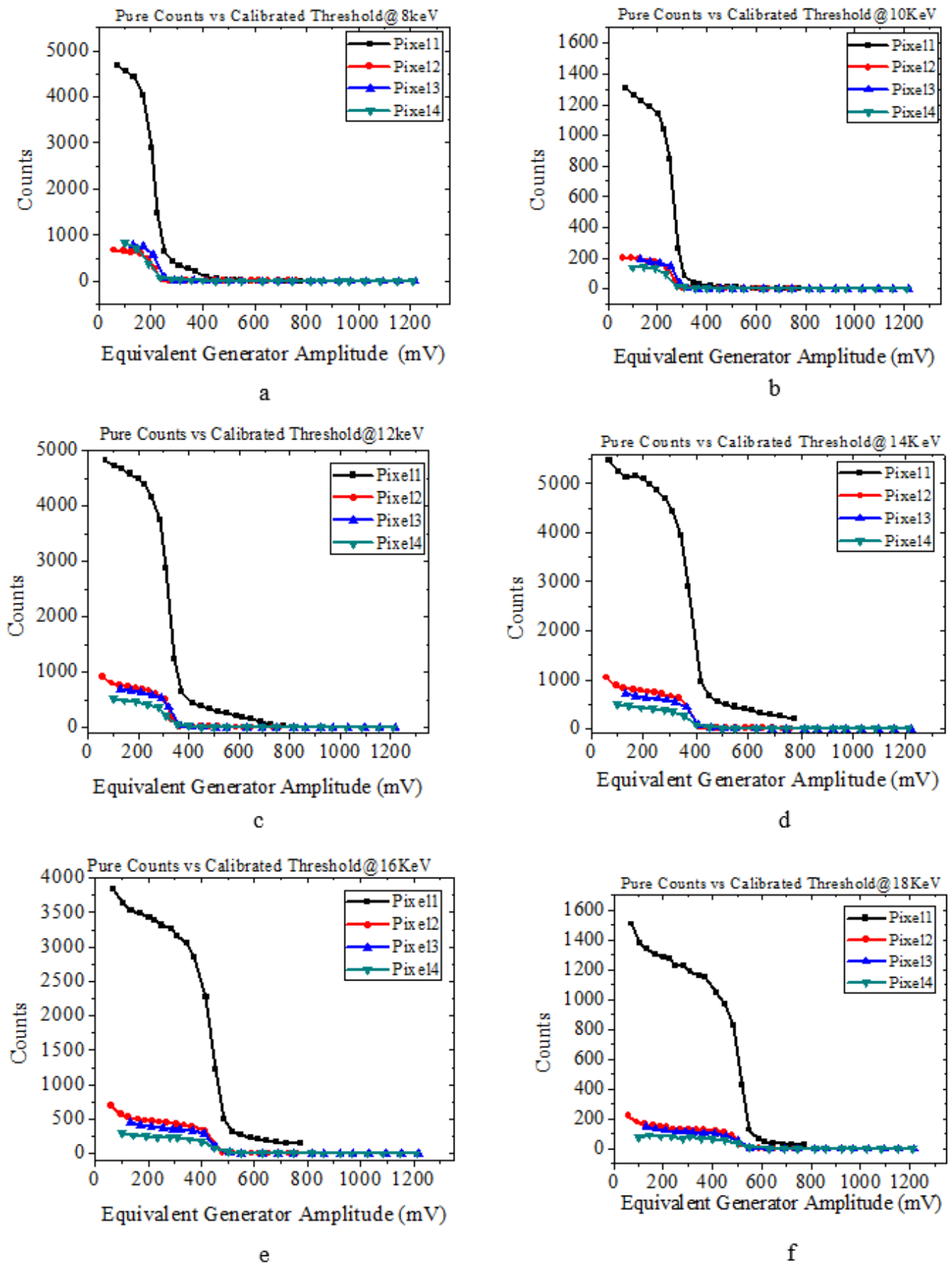

Fig.13. Four pixels count performance according to equivalent amplitude

at the X-ray energy of (a) 8*KeV* (b) 10*KeV* (c) 12*KeV* (d) *14KeV* (e) 16*KeV* (f) 18*KeV* for the different energies of X-ray, the sharp decreased points of the equivalent generator amplitudes are different.

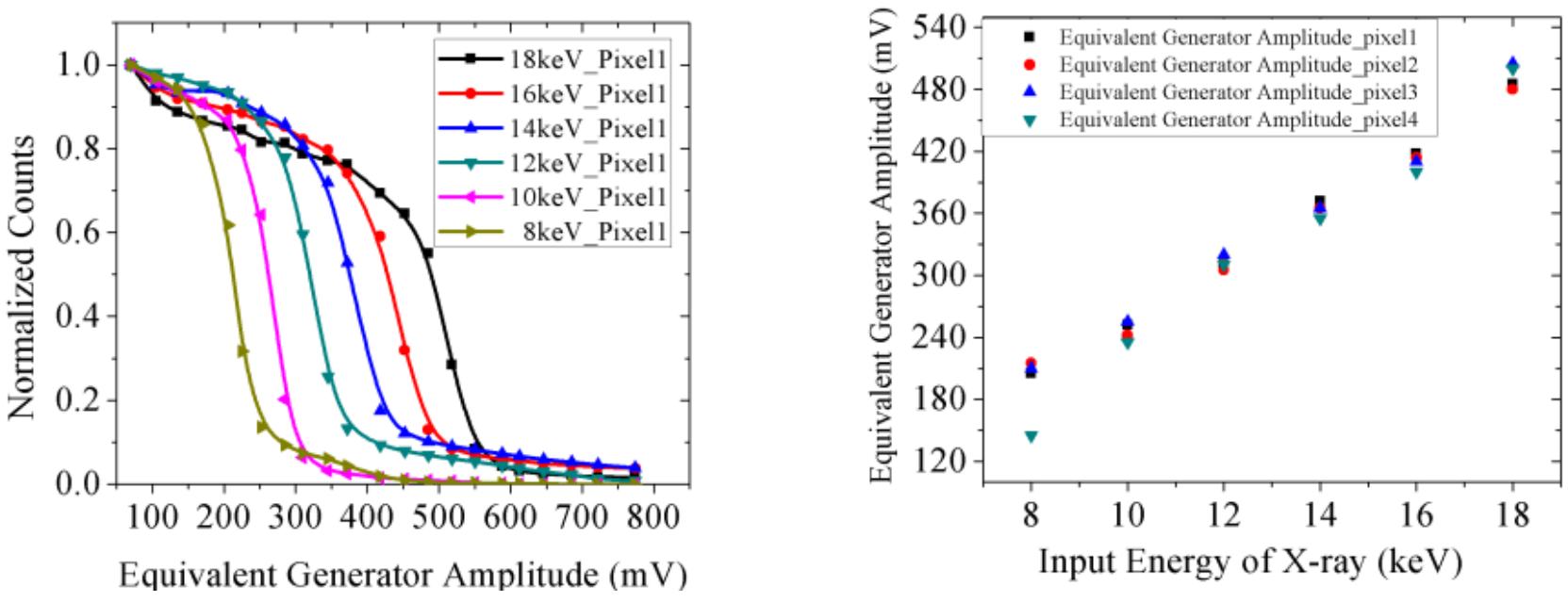

(a) (b)

Fig.14. (a) The normalized counts of pixel 1 at different energies, (b) the equivalent generator and the input X-ray energy show the linear relation.

The energy resolution of pixel sensor at different energy points is vary as sketched in figure 15. The differential of S-curve (figure 14(a)) is the energy spectrum and the energy resolution were calculate by the formula $Energy\ Resolution = \frac{FWHM}{Energy}$. From figure 15, we can see the higher energy value the better energy resolution. The pixels were wire bonded to the ASIC circuit, so the stray capacitance would be larger than that bonded via indium ball, and energy resolution could be gained better by indium bonding.

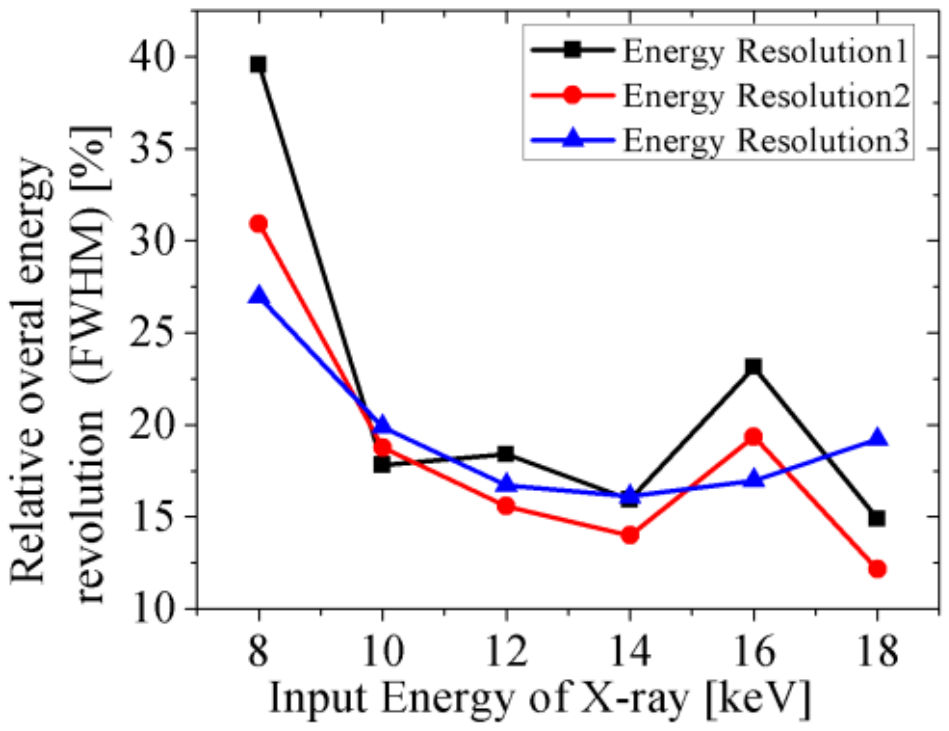

Fig.15. The energy resolution of three pixels at different energies

## 5 Summary and Outlook

In this paper, pixel sensors designed for the hybrid pixel single-photon-counting detector, for the detection of synchrotron X-ray detection have been studied. The silicon pixel sensors are designed for BAPS and will be firstly used in biomacromolecule structure elucidation。 Every real sensor is designed with 72×104 pixels with the pitch of 150μm and the sensor would bump bond to 8 ASICs. The detector will be finally constructed for large area with integrating the chip (consisting of 1 sensor and 8 ASICS) together. In the present work, we just bond the sensor (8×8 pixels for the sensors' performance) with wires to the ASIC and this would lead to larger stray capacitance which would enlarge the noise of detector system. The pixels sensor will be connected to the ASIC via bump bonding (indium balls packing), and the performance of the pixels sensor and ASIC will get improved especially about the noise and the energy resolution.

*This work was supported by Prefabrication research of Beijing Advanced Photon Source (R&D for BAPS).*